%% file: ms.tex
\documentclass[sigconf]{acmart}

\usepackage{graphicx}
\usepackage{subfiles}
\usepackage{todonotes}
\usepackage{multirow,tabularx}
\usepackage{booktabs}
\PassOptionsToPackage{hyphens}{url}\usepackage{hyperref}
\usepackage{breakurl}

\usepackage{dsfont}
\usepackage{amsmath}

\usepackage{mathtools}

\usepackage{caption}
\usepackage{subcaption}

\usepackage{xcolor}

\AtBeginDocument{%
  \providecommand\BibTeX{{%
    \normalfont B\kern-0.5em{\scshape i\kern-0.25em b}\kern-0.8em\TeX}}}

\copyrightyear{2021}
\acmYear{2021}
\setcopyright{acmlicensed}\acmConference[DocEng '21]{ACM Symposium on Document Engineering 2021}{August 24--27, 2021}{Limerick, Ireland}
\acmBooktitle{ACM Symposium on Document Engineering 2021 (DocEng '21), August 24--27, 2021, Limerick, Ireland}
\acmPrice{15.00}
\acmDOI{10.1145/3469096.3469872}
\acmISBN{978-1-4503-8596-1/21/08}

\begin{document}

\title{On Minimizing Cost in Legal Document Review Workflows}

\author{Eugene Yang}
\affiliation{%
  \institution{IR Lab, Georgetown University}
  \city{Washington}
  \state{DC}
  \country{USA}
}
\email{eugene@ir.cs.georgetown.edu}

\author{David D. Lewis}
\affiliation{%
  \institution{Reveal Data}
   \city{Chicago}
   \state{IL}
  \country{USA}
}
\email{doceng2021paper@davelewis.com}

\author{Ophir Frieder}
\affiliation{%
  \institution{IR Lab, Georgetown University}
  \city{Washington}
  \state{DC}
  \country{USA}
}
\email{ophir@ir.cs.georgetown.edu}

\renewcommand{\shortauthors}{Yang, Lewis, and Frieder}

\begin{abstract}

Technology-assisted review~(TAR) refers to human-in-the-loop machine learning workflows for document review in legal discovery and other high recall review tasks. Attorneys and legal technologists have debated whether review should be a single iterative process (one-phase TAR workflows) or whether model training and review should be separate (two-phase TAR workflows), with implications for the choice of active learning algorithm. The relative cost of manual labeling for different purposes (training vs. review) and of different documents (positive vs. negative examples) is a key and neglected factor in this debate.  Using a novel cost dynamics analysis, we show analytically and empirically that these relative costs strongly impact whether a one-phase or two-phase workflow minimizes cost. We also show how category prevalence, classification task difficulty, and collection size impact the optimal choice not only of workflow type, but of active learning method and stopping point. 

\end{abstract}



\maketitle

\subfile{1-intro}

\subfile{2-background}

\subfile{3-cost-model}

\subfile{4-some-typical-cost-structures}

\subfile{5-exp}

\subfile{6-results}

\subfile{7-summary}

\bibliographystyle{ACM-Reference-Format}
\bibliography{ms}

\end{document}

%% file: 1-intro.tex
\section{Introduction}

Manual review of large document collections is a key task in the law and other applications. In the law, \textit{electronic discovery} (\textit{eDiscovery}) or \textit{electronic disclosure} (\textit{eDisclosure}) refers to a range of enterprise document review tasks including review for responsiveness in civil litigation~\cite{baron2016perspectives,harty2017discovery}, regulatory reviews \cite{nasuti2014shaping}, and internal investigations \cite{holton2009identifying}. Structurally similar tasks include systematic review in medicine (finding  published clinical trials of a treatment) \cite{wallace2010semi} and content monitoring for hate speech and harrassment in social media~\cite{ghasem2015machine,halevy2020preserving}.

\textit{Technology-assisted review} (\textit{TAR}) refers to iterative human-in-the-loop workflows where experts label documents for relevance to a task, and supervised learning of predictive models from those labeled documents is used to find the next batch of documents for review \cite{baron2016perspectives}. In machine learning, this is referred to as an \textit{active learning} workflow \cite{settles2009active}. Exploding volumes of documents have made TAR widely used in large legal matters \cite{pace2012money}.

Low cost is not the only objective. TAR reviews often must meet an effectiveness target, typically on recall (the fraction of the relevant documents were found) \cite{lewis2016defining}.  A party in a litigation may, for example, agree to over at least 80\% of documents responsive to a request for production. 

TAR workflows vary in whether they distinguish (e.g. by using different personnel) between coding documents for training predictive models, and coding them to accomplish the review task itself. In a one-phase TAR workflow all reviewed documents are also used for training, while in a two-phase TAR workflow there are separate training and review phases. There is substantial debate, particularly in the law, over which workflow  to use (Section \ref{sec:back:controversy}).

A key but neglected factor in choosing among TAR workflows is that costs differ from reviewer to reviewer, and even from document to document. Our contributions in this paper are (1) a mathematical model that predicts how these varying costs affect total workflow cost (Section \ref{sec:cost-analysis}); (2) a new visualization tool, the {\em cost dynamics graph}, which displays the landscape of total costs encountered by an evolving workflow (Section \ref{sec:cost-structures}); and (3) an empirical study that tests the predictions of our mathematical model (Section \ref{sec:results}). We confirm our model's prediction that one-phase workflows are optimal under a mathematically narrow (but practically common) set of conditions. The results also provide new insights into choice of workflows, active learning methods, and stopping rules.

%% file: 2-background.tex
\section{TAR Workflows in the Law}

TAR workflows are widely applicable, but have seen their most extensive use in the law. 
The first analytics technology applied to finding documents relevant in civil litigation was Boolean text search, originally applied to manually keypunched abstracts of documents \cite{blair1985evaluation}. A Boolean query, typically a disjunction of words, phrases, and proximity operators referred to as a {\em keyword list}, is used to select a subset of collected documents for review \cite{baron2007sedona, oard2013information}. It thus serves as a binary (yes/no) \textit{classifier} \cite{lewis2016defining}. 
While some documents might be examined while creating the keyword list, most documents reviewed by the workflow are those that the query retrieves, i.e., those on which the classifier makes a positive prediction. We refer to a workflow where a classifier creation phase (phase one) is followed by a distinct document review phase (phase two) as a \textit{two-phase TAR workflow}.

As documents became increasingly digital, additional  technologies were applied.  Grouping technologies such as duplicate detection, near-duplicate detection, email threading, and document clustering allowed related documents to be reviewed together and/or decisions propagated to group members \cite{simek2009technology,joshi2011auto}. Text retrieval methods such as query reformulation, statistical ranked retrieval, and latent indexing became used in the law under the heading of ``concept search''~\cite{oard2010evaluation, oard2013information, laplanche2004concept}.

As these technologies became available, they were often used interactively. A user might examine a 
graphical display 
of document clusters, each having a summary description, and label each cluster as relevant or not relevant.~\footnote{\url{https://chrisdaleoxford.com/2008/09/21/attenex-round-every-corner}}  They might run multiple
ranked retrieval 
searches, and label one or many documents at the top of a ranking as relevant or not relevant. Even the result of a Boolean query might be bulk-labeled as relevant. 

The net result of such interactions is that a review decision has been made (explicitly or by a default assumption of non-relevance) for every document in the collection.  This user-created classification may be viewed as definitive (the interactive reviewers are the final reviewers)
or only tentative. In any case, while classifiers (e.g. Boolean queries) may have been created along the way, the final classification is the cumulative result of many user actions during the review, and does not correspond to any single final query or classifier.  We refer to workflows of this sort as \textit{one-phase TAR workflows}. 

\subsection{Supervised Learning in TAR Workflows} 

Supervised learning of predictive models from labeled documents (often referred to as \textit{predictive coding} in the law) began to be used in civil litigation in the mid-2000s \cite{baron2007sedona}, as well as attracting research attention \cite{tomlinson2007overview}. 
The first United States federal case explicitly approving its use in review for responsiveness (producing documents requested by adversaries) appeared in 2012.\footnote{Da Silva Moore v. Publicis Groupe (Da Silva Moore 17), No. 11 Civ. 1279(ALC)(AJP), 2012 WL 607412.  (S.D.N.Y. Feb. 24, 2012} Cases followed in Ireland\footnote{Irish Bank Resolution Corp. v. Quinn [2015] IEHC 175 (H. Ct.) (Ir).}, Australia\footnote{McConnell Dowell Constructors (Aust) Pty Ltd v Santam Ltd \& Ors (No.1) [2016] VSC 734}, 
England\footnote{Pyrrho Inv. Ltd. v. MWB Bus. Exch., Ltd., [2016] EWHC 256 (Ch) [1] (Eng.)}, and other jurisdictions.

An early focus 
was replacing 
Boolean queries in two-phase workflows with binary text classifiers produced by supervised learning \cite{oneill2009disco, peck2011search, baron2013cooperation}. Replacing manual query writing with supervised learning made the phases of a two-phase review more similar. Both now involved labeling documents, with the difference being only in whether the primary purpose of labeling was training or the final review dispensation.

Unlike Boolean queries, classifiers produced by supervised learning algorithms typically assign a numeric score to every document. That made them easy to use not just in a two-phase culling workflow, but also for prioritizing documents in an interactive one-phase workflow. Losey first proposed one-phase workflows where supervised learning was one of several text analytics tools used\footnote{\url{https://e-discoveryteam.com/2012/07/01/day-one-of-a-predictive-coding-narrative-searching-for-relevance-in-the-ashes-of-enron/}}, in the spirit of Bates' berrypicking formulation\footnote{\url{https://e-discoveryteam.com/2013/04/21/reinventing-the-wheel-my-discovery-of-scientific-support-for-hybrid-multimodal-search/}} of information access \cite{bates1989design}. Tredennick\footnote{\url{https://web.archive.org/web/20140420124327/http://www.catalystsecure.com/blog/2013/11/tar-2-0-continuous-ranking-is-one-bite-at-the-apple-really-enough/}}, Cormack \& Grossman \cite{cormack2009machine,cormack2014evaluation}, and others proposed one-phase workflows using supervised learning only or predominantly.

\subsection{Controversy in the Law} 
\label{sec:back:controversy}

There is currently intense debate in the legal world over workflow design. At one extreme, some commentators assert two-phase culling workflows are ``TAR 1.0'' \cite{tredennick2015tar} or "simple" active learning~\cite{cormack2014evaluation}, and promote one-phase  workflows under headings such as ``TAR 2.0'' \cite{tredennick2015tar}, ``TAR 3.0''\footnote{\url{https://www.law.com/legaltechnews/2020/11/19/tar-3-0-expectations-for-modern-review-technology/}}, ``Hybrid Multimodal Predictive Coding 4.0''\footnote{\url{https://e-discoveryteam.com/2018/10/08/do-tar-the-right-way-with-hybrid-multimodal-predictive-coding-4-0/}}, and so on.

Cormack \& Grossman have asserted the superiority of a one-phase using relevance feedback \cite{rocchio1965relevance} (training on top-ranked documents) workflow under the trademarked terms Continuous Active Learning (TM)
\footnote{United States Trademark 86/634255}
and CAL (TM)
\footnote{United States Trademark 86/634265}
in patent filings~\cite{cormack2020systems} and scholarly \cite{cormack2016scalability} work. 
Members of the judiciary have weighed in.\footnote{US Federal Judge Andrew Peck stated in 2015: ``If the TAR methodology uses `continuous active learning' (CAL) (as opposed to simple passive learning (SPL) or simple active learning (SAL)), the contents of the seed set is much less significant.'' \textit{Rio Tinto PLC v. Vale S.A., 306 F.R.D. 125, 128-29 (S.D.N.Y. 2015)}.} 

On the other hand, two-phase workflows based on Boolean text querying,  metadata querying, supervised learning, and other binary classifier formation methods continue to be widely used.\footnote{Court orders involving two-phase workflows in recent US cases include \textit{In Re Broiler Antitrust (N.D. Ill., Jan 3, 2018}, \textit{City of Rockford v. Mallinckrodt ARD Inc (N.D. Ill. Aug. 7, 2018)}, and \textit{Livingston v. City of Chicago, No. 16 CV 10156 (N.D. Ill. Sep. 3, 2020)}}  
The US Department of Justice Antitrust Division's model agreement for use of supervised learning assumes a two-phase culling workflow with \textit{no} human review of classifier responsiveness decisions in phase two.~\footnote{\url{https://www.justice.gov/file/1096096/download}}
Scholars have felt the need to push back and argue for attorney oversight of phase two machine learning decisions.~\cite{keeling2020humans}.

\subsection{Review Cost in TAR}
\label{subsec:review-cost-background}

Our view is that one-phase and two-phase workflows both have their place, and that the per-document cost structure of review is a neglected factor in choosing between them. 

To take an extreme example, under the DOJ order mentioned above, no human review for responsiveness occurs in phase two. Per-document costs in phase two are not zero (there is usually at least bulk screening for attorney-client privilege), but are lower than for labeling training data in phase one. A two-phase review is not only required by the order, but is economically advantageous.

More generally, different reviewers may be used in the two phases. The label assigned to a training document  affects the fate of many other documents, so review for training is sometimes viewed as requiring senior attorneys (e.g., associates billing at up to US \$1000 per hour\footnote{\url{https://www.law.com/americanlawyer/2020/05/22/associate-billing-rates-surpass-1k-as-firms-snap-up-bankruptcy-work/}}). Conversely, review decisions not used for training are often done by contract attorneys (perhaps billed out at US \$40 per hour\footnote{\url{https://www.theposselist.com/2018/06/18/tales-from-the-trenches-the-explosion-of-e-discovery-document-review-projects-in-d-c-and-nyc/}}).

A second factor is the relative cost of different types of documents. If most nonrelevant documents are obviously off-topic, this might require little reading (e.g., just an email subject line) to determine that.
Conversely, relevant documents may require full reading to make a decision.  Little published data on this are available, but \citet{mcdonald2020accuracy} found that coming to a classification decision for sensitive documents took longer than for nonsensitive documents. 
Further, responsive documents may require additional decisions, such as  privilege review \cite{oard2018jointly}.

TAR research has largely ignored this cost asymmetry. Systems have been evaluating using generic information retrieval metrics, or cost-based measures that treat all costs as equal \cite{tomlinson2007overview,grossman2010technology,oard2010evaluation,bagdouri2013towards,roegiest2015trec,cormack2016engineering,yang2017icail,li2020stop}. A rare exception is MINECORE, a proposed two-phase workflow for combined responsiveness and privilege review \cite{oard2018jointly}. MINECORE assumes, however, that multiple classifiers are used as well.

The CLEF 2017 eHealth Technology-Assisted Review proposed a ``cost-effective'' systematic review sub-task, where one evaluation metric took into account whether a reviewer examined just the abstract and title of a document, or read the full document \cite{clef2017ehealth-tar}. The CLEF 2018 eHealth Technology-Assisted Review web page proposed two such measures\footnote{\url{https://sites.google.com/view/clef-ehealth-2018/task-2-technologically-assisted-reviews-in-empirical-medicine}}. Unfortunately, it is not clear these measures were actually used: neither track overview discusses results on them  \cite{clef2017ehealth-tar,clef2018ehealth-tar}.

%% file: 3-cost-model.tex
\section{A Framework for TAR Costs}
\label{sec:cost-analysis}

The goal of most TAR workflows is to hit an effectiveness target (e.g., 0.80 recall) while minimizing cost. Choices such as workflow style, sampling strategy, and stopping point must be made
with an eye toward this goal. We propose that a finer-grained examination of costs can inform these decisions.

\subsection{An Idealized Cost Model}
\label{sec:cost-analysis-ideal}

\begin{figure*}
    \centering
    \includegraphics[width=\linewidth]{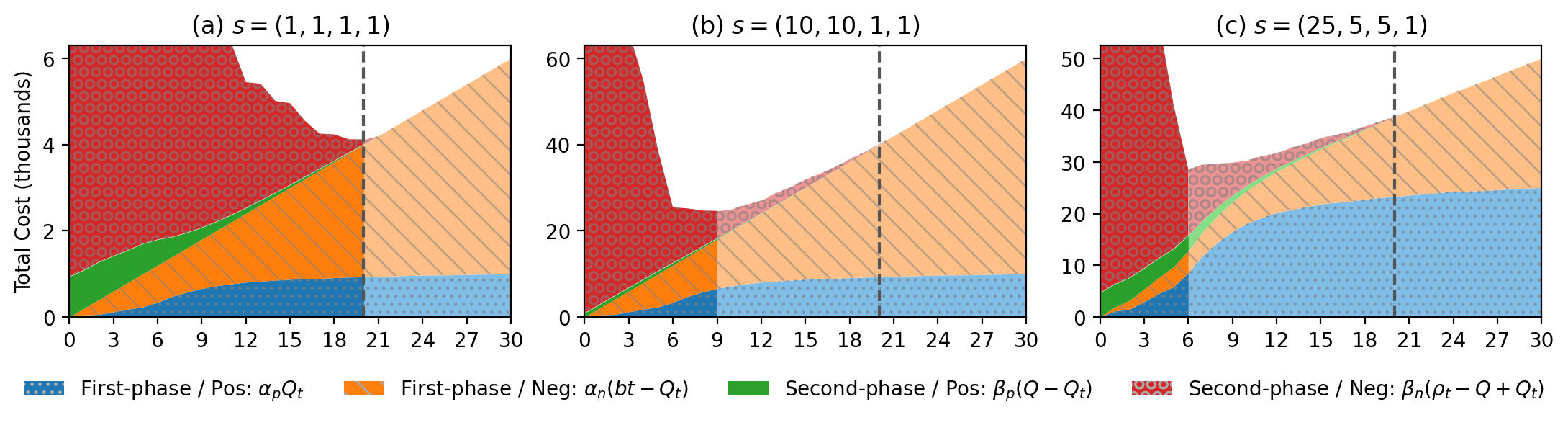}
    \vspace{-2.5em}
    \caption{Cost dynamics graphs for category \texttt{GENV} on 20\% RCV1-v2 collection. We assume relevance feedback with a 0.80 recall target under three cost structures. See Section~\ref{sec:exp} for experimental details. The x-axis shows number of iterations of active learning. Colored/shaded areas represent the four costs in Equation~\ref{eq:cost-by-review-type}, such that the height of the filled area equals total review cost when stopping at that iteration. Iterations after the minimum cost iteration are more lightly shaded. The dashed line at iteration 20 indicates when a one-phase review achieves 0.80 recall.}
    \label{fig:area-example}
    \vspace{-0.5em}
\end{figure*}

For the purposes of this study, we define a \textit{TAR cost structure} to be a four-tuple $s = (\alpha_p, \alpha_n, \beta_p, \beta_n)$. Here $\alpha$ and $\beta$ represent the cost of reviewing one document during the first and second phase respectively; the subscripts $p$ and $n$ indicate the cost for reviewing positive (e.g., relevant or responsive) and negative (e.g., nonrelevant or nonresponsive) documents, respectively. 

Assume a one-phase TAR workflow with uniform training batches of size $b$, stopping after $t$ iterations with $Q_t$ positive documents found. The cost incurred is $\alpha_{p}Q_t + \alpha_{n}(bt - Q_{t})$.  Stopping at iteration $t$ might not meet the recall goal, however, so we also need a \textit{failure cost} to account for remediation. 

For analysis purposes, we propose defining the failure cost for a one-phase review to be the cost of continuing on to an \textit{optimal} two-phase review. That is, we assume that the model trained on documents labeled during the $t$ iterations before stopping is used to rank the unreviewed documents, and those documents are reviewed in order until the recall target is hit. 
 
If a collection contains R positive documents, and the recall goal is $g$ (e.g., 0.8), then $Q = \lceil gR \rceil$ is the minimum number of documents that must be reviewed to reach the recall target. A one-phase review has a deficit of $Q - Q_t$ positive documents when $Q > Q_t$.  Let $\rho_t$ be the minimum number of documents that must be examined from the top of the ranking of unreviewed documents to find an additional $Q - Q_t$ positive documents.  Note that $\rho_t$ is reduced both by having found more positive documents in phase one, and by having trained a more effective model at the end of phase one to use in phase two.  

Given the above, we can define the phase two cost (failure penalty) to be $\beta_p (Q - Q_t) + \beta_n (\rho_t - Q + Q_t)$. The total cost, $Cost_s(t)$, of a one phase review with this failure penalty is, for $Q_t < Q$:  
\begin{equation}
   \alpha_p Q_t + \alpha_n (bt - Q_t) 
    + I[Q_t < Q] \left(
                  \beta_p (Q - Q_t) 
                  + \beta_n (\rho_t - Q + Q_t)
                 \right)
    \label{eq:cost-by-review-type}
\end{equation}
where $I[Q_t < Q]$ is 1 if $Q$ documents were not found in the first $t$ iterations, and 0 otherwise.  

We then define the cost of a two-phase review stopping at $t$ similarly, i.e., we assume the second phase is conducted optimally, reviewing the same $\rho_t$ documents that a failed one-phase review would. Indeed, in this framework, a one-phase review is simply a two-phase review that stops at a point where the second phase is not necessary.  

Our idealized definition of failure cost has several advantages.  It is a deterministic, easily computed value defined for all stopping points and independent of the stopping rule used. More importantly, it puts one-phase and two-phase reviews on the same footing. When statistical guarantees of effectiveness are needed, two-phase reviews often use a labeled random sample to choose a second-phase cutoff. Proponents of one-phase reviews frequently point to the cost of labeling this random sample, while ignoring the fact that holding a one-phase review to the same statistical standard would incur a similar cost.

\subsection{Cost Dynamics Graphs}

Our framework allows costs to be compared equitably not just between workflow types, but across iterations within a single workflow. As a visualization of how workflow cost evolves over iterations, we propose using a \textit{cost dynamics graph} that plots our total cost measure at each possible stopping point, while breaking out components of the cost separately. 

Figure~\ref{fig:area-example} provides an example. Total review cost after each iteration is separated into first phase positives (blue w/ dots), first phase negatives (orange w/ strokes), second phase positives (uniform green), and second phase negatives (red w/ circles). Costs here are for the same TAR workflow execution, but plotted under three different cost structures (Section \ref{sec:cost-structures}). The workflow is carried out for 30 iterations of relevance feedback with a training batch size of 200 (details in Section \ref{sec:exp}).   

With a uniform cost structure (a), minimum cost (lowest height of the total shaded area) is reached at iteration 20 where (almost) no second phase review is needed: a one-phase review is basically optimal\footnote{While difficult to see in (a), there actually are a small number of documents, a fraction of one batch, reviewed in phase two.}.  For cost structure (b), where phase one is ten times as expensive as phase two, stopping much sooner (at iteration 9) is optimal, with an 41\% cost reduction over the essentially one phase review ending at iteration 20. For cost structure (c), where both phase one and positive documents are five times more costly than their counterparts, stopping at iteration 6 is optimal, with 28\% of cost reduction over stopping at iteration 20.

\subsection{Fixed Versus Variable Costs}

Equation~\ref{eq:cost-by-review-type} is awkward to use directly for reasoning about workflow costs. If we rewrite to collect terms in $Q$, $Q_t$, and $\rho_t$ we get, for $Q_t \le Q$, that $Cost_s(t)$ is  
\begin{equation}
    (\alpha_p - \alpha_n - \beta_p + \beta_n) Q_t + \alpha_n bt + \beta_n\rho_t + (\beta_p - \beta_n) Q,
    \label{eq:cost-by-components}
\end{equation}
while for $Q_t \ge Q$ it is simply 
\begin{equation}
    (\alpha_p - \alpha_n) Q_t + \alpha_n bt. 
    \label{eq:cost-by-components-nosecond}
\end{equation}

For a given target $Q$ and cost structure $s$, the fourth term, $(\beta_p - \beta_n) Q$, is a fixed cost. The remaining terms are variable costs that reflect characteristics of the TAR approach.  The first term depends on the number of positives found by iteration $t$,  while the second term, $\alpha_n bt$, is simply linear in the number of iterations $t$. The third term, $\beta_n\rho_t$, depends on both the undone work at iteration $t$ and the quality of the predictive model formed by then. 
In the next section, we use this decomposition of costs to predict TAR behavior under typical cost structures.

%% file: 4-some-typical-cost-structures.tex
\section{Some TAR Cost Structures}
\label{sec:cost-structures}

We now examine typical cost structures and their implications for workflow style and active learning method.  

\subsection{Uniform Cost} 
\label{sec:cost-model:uniform}

For many review projects, it is reasonable to assume all review costs are roughly equal. If so, we can without loss of generality assume $s = (1,1,1,1)$, and thus total cost is  
\begin{equation}
    Cost_{uniform}(t) = bt + \rho_t
    \label{eq:cost-uniform}
\end{equation}
where $\rho_t = 0$ if $Q_t \ge Q$.

Suppose that we conduct a relevance feedback review under this cost structure, and make the following plausible assumptions: (1) more training data lead to better models, and (2) training batches with more positives lead to better models than training batches with fewer positives. Neither is uniformly true, and the second is often false early in training \cite{lewis1995sequential}.  Later in a TAR review, however, both tend to be reasonable assumptions. We also assume that batch size $b$ is small to neglect any benefit that a two-phase review would get from reviewing less than a full batch. 

We formalize Assumption 1 by positing that, for any set of documents and any cutoff $k$, the top $k$ documents from that set have precision equal or higher when ranked by model $M_{t+1}$ than by model $M_t$, where $M_t$ is the model trained on $t$ batches. 

Consider the decision after iteration $t$ of whether to continue a one-phase review or switch to phase two.  Let $U_t$ be the set of unreviewed documents after $t$ iterations. Whether we continue a one-phase review or switch to a two-phase review, the next $b$ documents reviewed will be $B_{t+1}$, the top $b$ documents from a ranking of $U_t$ induced by $M_t$. With a uniform structure, the cost is the same whether $B_{t+1}$ is reviewed in phase one or phase two.    

In a one-phase review, $B_{t+1}$ will be added to the training set. Model $M_{t+1}$ will be produced and used to rank $U_t \setminus B_{t+1}$.  The next $b$ documents, $B_{t+2}$ will be drawn from $M_{t+1}$'s ranking of $U_t \setminus B_{t+1}$.  In a two-phase review $B_{t+1}$ will also be reviewed, but not used for training.  The next $b$ documents ($B'_{t+2}$) will be drawn instead from $M_{t}$'s ranking of $U_t \setminus B_{t+1}$.  

The cost of reviewing $B_{t+2}$ or $B'_{t+2}$ is the same. Under our assumptions, however, a one phase review has both immediately found more positive documents ($B_{t+2}$ will have more on average than $B'_{t+2}$), and will have a better  model ($M_{t+1}$ instead of $M_{t}$) to find future documents. We therefore should not transition to phase two. Since the same analysis applies at every iteration,\textbf{ we predict that a one-phase review is optimal in this setting.} 

This analysis, while not a proof, suggests why one-phase relevance feedback workflows have shown good results with uniform costs~\cite{cormack2014evaluation,cormack2016engineering}. Figure~\ref{fig:area-example}(a) shows an example of cost dynamics with a uniform cost structure.

\subsection{Expensive Training}
\label{subsec:expensive-training}

When more senior attorneys are required for training, per document review costs in phase one may be a factor of 10 or more higher than in phase two (Section~\ref{subsec:review-cost-background}).  A similar asymmetry can occur in systematic review in medicine~\cite{clef2017ehealth-tar,clef2018ehealth-tar}.

Such cost structures have the form $(\alpha, \alpha, \beta, \beta)$ with total cost 
\begin{equation}
    Cost_{phased}(t) = \alpha bt + \beta\rho_t
    \label{eq:cost-twophase}
\end{equation}
where $\rho_t = 0$ if $Q_t \ge Q$.
We usually have $\alpha > \beta$, so a one phase review is optimal only if every training batch of size $b$ improves the classifier enough to eliminate $b\alpha / \beta$ documents from the competing phase two review.  Since learning curves show diminishing returns with training data \cite{kolachina2012prediction}, \textbf{we predict a two phase review will usually be optimal in this setting.} Whether relevance feedback or some other active learning method will dominate is less clear. Scenario (b) in Figure~\ref{fig:area-example} shows an example.

\subsection{Expensive Positives}
\label{sec:cost-model:poscost}

In this section, we consider several scenarios where positive examples are more expensive to review than negative examples.

\subsubsection{Additive Cost for Positives.}
\label{subsubsec:additivepositives}

In the law, positive documents may require review for factors (e.g.,  attorney-client privilege) not applicable to negative documents. One model for this is a fixed additional cost $v$ for each positive document, regardless of the phase in which it is found, i.e., structure $(\alpha+v, \alpha, \beta+v, \beta)$. The cost function when $Q_t \le Q $ is
\begin{align}
    Cost(t) & = \left((\alpha+v) - \alpha - (\beta+v) + \beta \right) Q_t \nonumber \\ 
            &\phantom{==} + \alpha bt + \beta\rho_t + \left( (\beta+v) - \beta \right)Q \nonumber\\
            & =  \alpha bt + \beta\rho_t + vQ
    \label{eq:cost-additive-pos}
\end{align}
and when $Q_t > Q$, becomes $\alpha bt + vQ + v(Q_t - Q)$.

This is the same as Equation~\ref{eq:cost-twophase} (or, if $\alpha = \beta$, Equation~\ref{eq:cost-uniform}), plus a fixed cost $vQ$, plus an additional penalty for overshooting $Q$. Since overshooting $Q$ is never optimal (ignoring issues of finite batch size), the optimal stopping point actually is the same as for cost structure $(\alpha, \alpha, \beta, \beta)$, so this scenario is not unique.

\subsubsection{Multiplicative Cost for Positives.}

Another possibility is that positives take more time, and that extra time incurs cost at the usual rate for each review phase. This implies a cost structure $(u\alpha, \alpha, u\beta, \beta)$ where $u>1$ is the multiplicative surcharge for the positive documents. The cost function for $Q_t \le Q$, is written as:  
\begin{align}
    Cost(t) &= (u\alpha - \alpha - u\beta + \beta) Q_t + \alpha bt + \beta\rho_t + (u\beta - \beta)Q \nonumber \\
            &= (u-1)(\alpha-\beta)Q_t + \alpha bt + \beta\rho_t + (u-1)\beta Q
\end{align}
and for $Q_t > Q$ is $(u-1)\alpha Q_t + \alpha bt$.

When $\alpha = \beta$, then $(u\alpha, \alpha, u\beta, \beta)$ is equivalent to the additive form $(\alpha+v, \alpha, \alpha+v, \alpha)$ with $v=(u-1)\alpha > 0$ (Section~\ref{subsubsec:additivepositives}). 
Multiplicative cost for positives is therefore only a unique scenario when $\alpha \neq \beta$. Typically this will be $\alpha > \beta$ (phase one is more expensive), so $(u-1)(\alpha - \beta)$ is nonnegative: there is a penalty for positive documents being found in phase one. 

That favors active learning strategies that find only those positives most useful for training. \textbf{We predict a two-phase review using a classifier-focused active learning strategy such as uncertainty sampling \cite{lewis1994sequential} will outperform both one-phase and two-phase reviews using relevance feedback.} Figure~\ref{fig:area-example}(c) displays such a cost structure, with minimum cost occurring at iteration 6. 

\subsubsection{Elite Phase One Review.}

Determining privilege can be more subtle legally than determining responsiveness. If elite reviewers are used during phase one, they may be able to incorporate privilege determination into their review at no additional cost. In contrast, responsive documents discovered during phase two may require calling on an elite reviewer to make the privilege decision.  We model this with the cost structure $(\alpha, \alpha, \beta_p, \beta_n)$ where $\alpha \ge \beta_p > \beta_n$. Total cost when $Q_t \le Q$ is, 
\begin{align}
    Cost(t) = -(\beta_p - \beta_n) Q_t + \alpha bt + \beta_n \rho_t + (\beta_p - \beta_n)Q
\end{align}
and when $Q_t > Q$ is simply $\alpha bt$. 

Given $\beta_p > \beta_n$, the coefficient for $Q_t$ is negative, rewarding finding positive documents in phase one. This favors relevance feedback as the active learning strategy. However, phase two review is still cheaper, particularly for negatives, so as batch precision declines we expect a transition to phase two at some point will be optimal. \textbf{We predict a two-phase review using relevance feedback to be optimal.}

%% file: 5-exp.tex
\section{Methods}
\label{sec:exp}

Our experiments test the predictions of our cost model, with an emphasis on how workflow choice, cost structure, and task properties interact.  

\subsection{Dataset}
\label{sec:exp:dataset}

Experimental evaluations of active learning require large, completely labeled, document collections. We simulate TAR tasks on two collections---RCV1-v2~\cite{rcv1} and the Jeb Bush email collection~\cite{totalrecall2015}---widely used in TAR research~\cite{cormack2016engineering, oard2018jointly, yang2019sigir, yang2019icail, totalrecall2015, totalrecall2016}. RCV1-v2 contains 804,414 news articles coded for each of 658 categories. We run some tests on all documents, and some on a fixed 20\% random subset (160,882 documents), to study the effect of collection size. The Jeb Bush collection, after deduplication, consists of 274,124 emails to and from the governor of the US state Florida, coded for 45 categories developed in two TREC (Text REtrieval Conference) evaluations of TAR technology~\cite{totalrecall2015,totalrecall2016}.

To study the impact of task characteristics, we chose 5 random RCV1-v2 categories from each of 9 bins, based on three ranges of prevalence (class frequency) and three ranges of task difficulty~(Table~\ref{tab:rcv1-cate-counts}). Extremely low and high prevalence bins were omitted due the inability to keep the ratio of the bin boundaries small. Difficulty bins were based on the effectiveness of a logistic regression model trained on a random 25\% of the full RCV1-v2 and evaluated by R-precision on the remaining 75\%.  The Jeb Bush collection has too few categories for binning, so we simply used the 41 categories with 80 or more documents.

\begin{table}[t]
\caption{Number of categories in each of 15 bins on the full RCV1-v2 collection. Difficulty values are R-Precision when training on 25\% of collection and testing on 75\%.}
\label{tab:rcv1-cate-counts}
\vspace{-1em}
\begin{tabular}{l|rrr}
\toprule
					       &  \multicolumn{3}{c}{Difficulty} \\
					       &  Hard &Medium &  Easy\\   
Prevalence (by \# Pos Docs) &  (< 0.65) &  (0.85 - 0.65) &  (1.0 - 0.85) \\
\midrule
Too rare (< 500)	    &  204	&  56	&  10 \\
Rare (500 - 2,000)	    &  74 	&  56	&  26 \\
Medium (2,000 - 8,000)	&  47 	&  44	&  47 \\
Common (8,000 - 32,000)	&  9 	&  28	&  29 \\
Too common (> 32,000)	&  3 	&  11	&  14 \\
\bottomrule
\end{tabular}
\vspace{-1em}
\end{table}

\subsection{Implementation}
\label{sec:exp:implementation}

We implemented TAR workflows using \texttt{libact}~\cite{libact}, a Python framework for active learning. One-phase and two-phase workflows were run using each of two active learning algorithms: uncertainty sampling~\cite{lewis1994sequential} and relevance feedback~\cite{cormack2014evaluation,rocchio1965relevance}. We used these algorithms because of their prominence in TAR research \cite{cormack2014evaluation,cormack2015autonomy,yu2018finding,mcdonald2020active,mcdonald2020accuracy}, and because they make opposite choices with respect to the exploration / exploitation tradeoff in active learning \cite{lewis1994sequential,osugi2005balancing}. 

Supervised learning used the \texttt{scikit-learn} implementation of logistic regression, with an L2 regularization weight of 1.0 and the \texttt{CountVectorizer} as the tokenizer.\footnote{\url{https://scikit-learn.org/}} All words were used as features, with feature value equal to the BM25-style term frequency weight~\cite{robertson2004understanding,yang2019sigir}. 

\subsection{Evaluation}
\label{sec:exp:evaluation}

Our evaluation metric was our idealized total cost (Section~\ref{sec:cost-analysis}) to reach the recall target, which we set at 0.8. In applied settings, TAR workflows use heuristics and/or sample-based estimates to decide when to stop a one-phase review \cite{cormack2016engineering,saha2015batch-mode,callaghan2020statistical,li2020stop,anonymous-qbcb}. Similar methods are used in two-phase workflows to decide when to switch to phase two, and where to set the phase two cutoff. One goal of our experiments was to characterize the cost landscape within which such rules operate. 

We therefore continued all runs until a recall of at least 0.8 was reached, guaranteeing we include the optimal stopping point for both one-phase and two-phase workflows, and computed our total cost (and its components) at each iteration.

The first training batch in an active learning workflow typically includes one or more positive "seed" documents known to the user or found by a keyword search. To simulate this, Batch 0 of our active learning runs was a single seed document chosen at random from all positives for that category in the collection. For each category, we ran iterative active learning with 10 different seeds (to simulate varying user knowledge). The same seeds were used with both uncertainty sampling and relevance feedback. Subsequent batches were 200 documents (a typical batch size in legal review environments) chosen by active learning. This gave a total of 900 runs on full RCV1-v2, 900 runs on 20\% RCV1-v2, and 820 runs on Jeb Bush. A total of 541,733 predictive models were trained. Wall clock run time given the computational resources available at our institution was 3 weeks.

To study the overall benefit of a workflow, we computed the mean over a set of tasks (category/seed pairs) of the  \textit{relative cost reduction} achieved by using workflow A instead of workflow B. Specifically, we compute $1 - Cost_A(t_A) / Cost_B(t_B)$, where $t_A$ and $t_B$ are the optimal stopping iterations for A and B.  Values are thus in the range $(-\infty, 1.0)$.  

We tested statistical significance of the relative cost reduction using the two-sample Kolmogorov–Smirnov~(K-S) test~\cite{hodges1958significance} to avoid distributional assumptions.  A Bonferroni correction~\cite{bonferroni_correction} was applied for 84 tests in Table~\ref{tab:main-cost-comparison}  and 126 tests in Table~\ref{tab:breakdown-cost-of-2p}. (These counts include one cost structure originally studied, but then dropped from our presentation after concluding it was economically unrealistic.)

Cost structures were chosen to exemplify the non-redundant scenarios from Section~\ref{sec:cost-structures}: Uniform $(1,1,1,1)$; Expensive Training $(2,2,1,1)$ and $(10, 10, 1, 1)$; Expensive Training with Multiplicative Positives $(20, 10, 2, 1)$ and $(25, 5, 5, 1)$; and Elite Phase One Review  $(20,20,11,1)$.   As discussed in Section~\ref{sec:cost-structures}, other scenarios are redundant from the standpoint of optimal stopping iteration, and thus the comparison between one-phase and two-phase workflows.

%% file: 6-results.tex
\section{Results and Analysis}
\label{sec:results}

Our experiments tested predictions for how cost structure impacts the optimal choice of workflow style and active learning method, and examined the impact of task characteristics on this relationship.

\begin{table*}
\caption{Mean relative cost reduction resulting from using workflow A instead of workflow B under six cost structures, with negative values indicating workflow A is worse. \texttt{1P} and \texttt{2P} indicate one-phase and two-phase workflows; \texttt{Unc.} and \texttt{Rel.} indicate uncertainty sampling and relevance feedback. Values are means over 450 category/seed pairs for 100\% and 20\% RCV1-v2, and 330 for Jeb Bush. Italicized values are nonzero only due to a two-phase review stopping with the equivalent of a partial training batch. Values with a * indicate a statistically significant difference in distribution between A and B at 95\% confidence level using a two-sample Kolmogorov–Smirnov test with Bonferroni correction.}
\label{tab:main-cost-comparison}

\newcommand{\z}{\phantom{*}}
\newcolumntype{Y}{>{\raggedleft\arraybackslash}X}
\newcommand{\minitab}[2][l]{\begin{tabular}{@{}#1@{}}#2\end{tabular}}

\vspace{-0.5em}
\begin{tabularx}{0.95\textwidth}{l|l| YYY|YY|Y }
\toprule
Collection	& Workflow A vs. B  &  $(1,1,1,1)$ & $(2,2,1,1)$ & $(10,10,1,1)$ & $(20,10,2,1)$ & $(25,5,5,1)$ & $(20,20,11,1)$ \\
\midrule
\multirow{4}{*}{\minitab[l]{100\% RCV1-v2\\($N=804,414$)}}
    & \texttt{2P Unc. vs. 1P Rel.} & -0.0982\z &   0.0977* &   0.3969* &   0.4670* &  0.4160* &    0.2922* \\
    \cmidrule{2-8}
    & \texttt{2P Unc. vs. 2P Rel.} & -0.1340\z & -0.0446\z & -0.0591\z &   0.0524* &  0.1339* &  -0.0632\z \\
    & \texttt{2P Rel. vs. 1P Rel.} &  \textit{0.0237\z} &  0.1091\z &   0.3639* &   0.3862* &  0.3003* &    0.3263* \\
    & \texttt{1P Rel. vs. 1P Unc.} &   0.6098* &   0.6098* &   0.6098* &   0.5694* &  0.4878* &    0.6098* \\
\midrule
\multirow{4}{*}{\minitab[l]{Jeb Bush\\($N=274,124$)}}
    & \texttt{2P Unc. vs. 1P Rel.} &  -0.1476* & -0.0022\z &   0.2221* &   0.3001* &   0.3115* &    0.1391* \\
    \cmidrule{2-8}
    & \texttt{2P Unc. vs. 2P Rel.} &  -0.2602* &  -0.2049* &  -0.2374* & -0.1675\z & -0.0634\z &   -0.2668* \\
    & \texttt{2P Rel. vs. 1P Rel.} &  \textit{0.0803\z} &   0.1634* &   0.3767* &   0.4071* &   0.3559* &    0.3211* \\
    & \texttt{1P Rel. vs. 1P Unc.} &   0.4743* &   0.4743* &   0.4743* &   0.4304* &   0.3456* &    0.4743* \\
\midrule
\multirow{4}{*}{\minitab[l]{20\% RCV1-v2\\($N=160,882$)}}
    & \texttt{2P Unc. vs. 1P Rel.} & -0.2104\z & -0.0650\z &   0.2580* &   0.3472* &  0.3093* &    0.2460* \\
    \cmidrule{2-8}
    & \texttt{2P Unc. vs. 2P Rel.} & -0.3149\z &  -0.3171* &  -0.4697* &  -0.3255* &-0.1328\z &  -0.3741\z \\
    & \texttt{2P Rel. vs. 1P Rel.} &  \textit{0.0585\z} &  0.1628\z &   0.4665* &   0.4804* &  0.3676* &    0.4660* \\
    & \texttt{1P Rel. vs. 1P Unc.} &   0.4200* &   0.4200* &   0.4200* &   0.3863* &  0.3190* &    0.4200* \\
\bottomrule
\end{tabularx}
\end{table*}

\subsection{Costs, Workflows, and Active Learning}
\label{subsec:costflowactive}

As predicted by our cost model, Table~\ref{tab:main-cost-comparison} shows that a two-phase workflow has lower cost than a one-phase workflow for several asymmetric cost structures.  Claims that one-phase relevance feedback workflows are always superior (Section \ref{sec:back:controversy}) are simply incorrect.  

In contrast, and also as predicted, the one-phase workflow \textit{is} preferred for the uniform cost structure $(1,1,1,1)$, with a 10 to 20\% reduction in cost vs. the two-phase uncertainty sampling workflow. The situation is mixed for the slightly asymmetric structure $(2,2,1,1)$: a two-phase uncertainty workflow is better on 100\% RCV1-v2, but one-phase relevance feedback is better on the two smaller collections. 

Under any cost structure, the two-phase uncertainty workflow has a larger advantage or smaller disadvantage on full RCV1-v2 versus 20\% RCV1-v2. This emphasizes a neglected point in discussions of TAR workflows: the larger a data set, the more that costs incurred to improve classifier effectiveness are amortized over many documents.

Also neglected in eDiscovery discussions is the fact that relevance feedback can be used in two-phase workflows. Indeed, we find that relevance feedback dominates uncertainty sampling for two-phase workflows except for the Expensive Training with Multiplicative Positives scenarios $(20,10,2,1)$ and $(25,5,5,1)$ on full RCV1-v2. Providing more positive documents in training (which reviewers tend to prefer) can sometimes be a win/win situation. Section \ref{sec:results-task-charac} shows that the underlying story is more complex, when we focus on categories with particular properties. 

In Table~\ref{tab:main-cost-comparison} a two-phase relevance feedback workflow always dominates a one-phase relevance feedback workflow. However, for $(1,1,1,1)$ that is an artifact of finite batch size: we can usually  slightly reduce cost by replacing the last batch with an optimal cutoff in a second phase. The effect is small on the largest collection and would be negligible with smaller batch sizes.  

Finally, a one-phase workflow using uncertainty sampling always has poor effectiveness, since there is no second phase to pay back the expense of training a better classifier.

\subsection{Impact of Task Characteristics}
\label{sec:results-task-charac}

\begin{table*}[t]
\caption{Mean relative cost reduction when using uncertainty sampling rather than relevance feedback in a two-phase workflow (\texttt{2P Unc.} vs. \texttt{2P Rel.}) on full and 20\% RCV1-v2 collection. Table details are as in Table~\ref{tab:main-cost-comparison}.  Means are over 50 runs: 5 categories per bin (Section~\ref{sec:exp:dataset}) and 10 random seeds per category.}
\label{tab:breakdown-cost-of-2p}
\newcommand{\z}{\phantom{*}}
\newcolumntype{Y}{>{\raggedleft\arraybackslash}X}
\vspace{-0.5em}
\begin{tabularx}{0.9\textwidth}{l|l|l| YYY|YY|Y }
\toprule
Size &  Difficulty & Prevalence &   $(1,1,1,1)$ &    $(2,2,1,1)$ &  $(10,10,1,1)$ &  $(20,10,2,1)$ &   $(25,5,5,1)$ & $(20,20,11,1)$ \\
\midrule
\multirow{9}{*}{100\%}
&        & Common &  -0.0524\z &   0.1922* &   0.2089* &   0.2473* &   0.2343* &    0.0655\z \\
& Easy   & Medium &  -0.1029\z &   0.1052* &  -0.0188\z &   0.0898\z &   0.1780* &    0.0053\z \\
&        & Rare   &  -0.8323* &  -0.9031* &  -1.5450* &  -0.9403* &  -0.2893\z &   -1.0047* \\
\cmidrule{2-9}
&        & Common &   0.0200\z &   0.2136* &   0.4626* &   0.5027* &   0.4677* &    0.2427* \\
& Medium & Medium &   0.0003\z &   0.1127\z &   0.2005\z &   0.2703\z &   0.2918\z &    0.0829\z \\
&        & Rare   &  -0.1300\z &  -0.0701\z &  -0.0405\z &   0.0060\z &   0.0541\z &   -0.0815\z \\
\cmidrule{2-9}
&        & Common &   0.0326\z &   0.0829\z &   0.2248* &   0.2606* &   0.2425* &    0.1404\z \\
& Hard   & Medium &  -0.0283\z &  -0.0020\z &   0.0769\z &   0.1285* &   0.1439* &    0.0276\z \\
&        & Rare   &  -0.1131\z &  -0.1324\z &  -0.1014\z &  -0.0932\z &  -0.1182\z &   -0.0468\z \\
      
\midrule

\multirow{9}{*}{20\%}
&        & Common &  -0.2137* &  -0.1707\z &  -0.5981* &  -0.3383\z &  -0.0524\z &   -0.2425* \\
& Easy   & Medium &  -0.4012* &  -0.4491* &  -1.0240* &  -0.6919* &  -0.2362\z &   -0.5315* \\
&        & Rare   &  -1.4201* &  -1.5464* &  -1.9346* &  -1.5411* &  -0.9013* &   -1.6534* \\
\cmidrule{2-9}
&        & Common &  -0.0546\z &   0.0266\z &   0.0836\z &   0.1715\z &   0.2194* &   -0.0083\z \\
& Medium & Medium &  -0.2177\z &  -0.2168\z &  -0.2814\z &  -0.1987\z &  -0.0743\z &   -0.3272\z \\
&        & Rare   &  -0.3091\z &  -0.3028\z &  -0.3599\z &  -0.2605\z &  -0.1610\z &   -0.3476\z \\
\cmidrule{2-9}
&        & Common &  -0.0044\z &   0.0103\z &   0.1121\z &   0.1058\z &   0.1141\z &    0.0383\z \\
& Hard   & Medium &  -0.1150\z &  -0.1087\z &  -0.1304\z &  -0.0823\z &  -0.0043\z &   -0.1948\z \\
&        & Rare   &  -0.0985\z &  -0.0962\z &  -0.0950\z &  -0.0937\z &  -0.0990\z &   -0.0996\z \\
\bottomrule
\end{tabularx}
\end{table*}

\begin{figure*}
    \centering
    \includegraphics[width=0.95\linewidth]{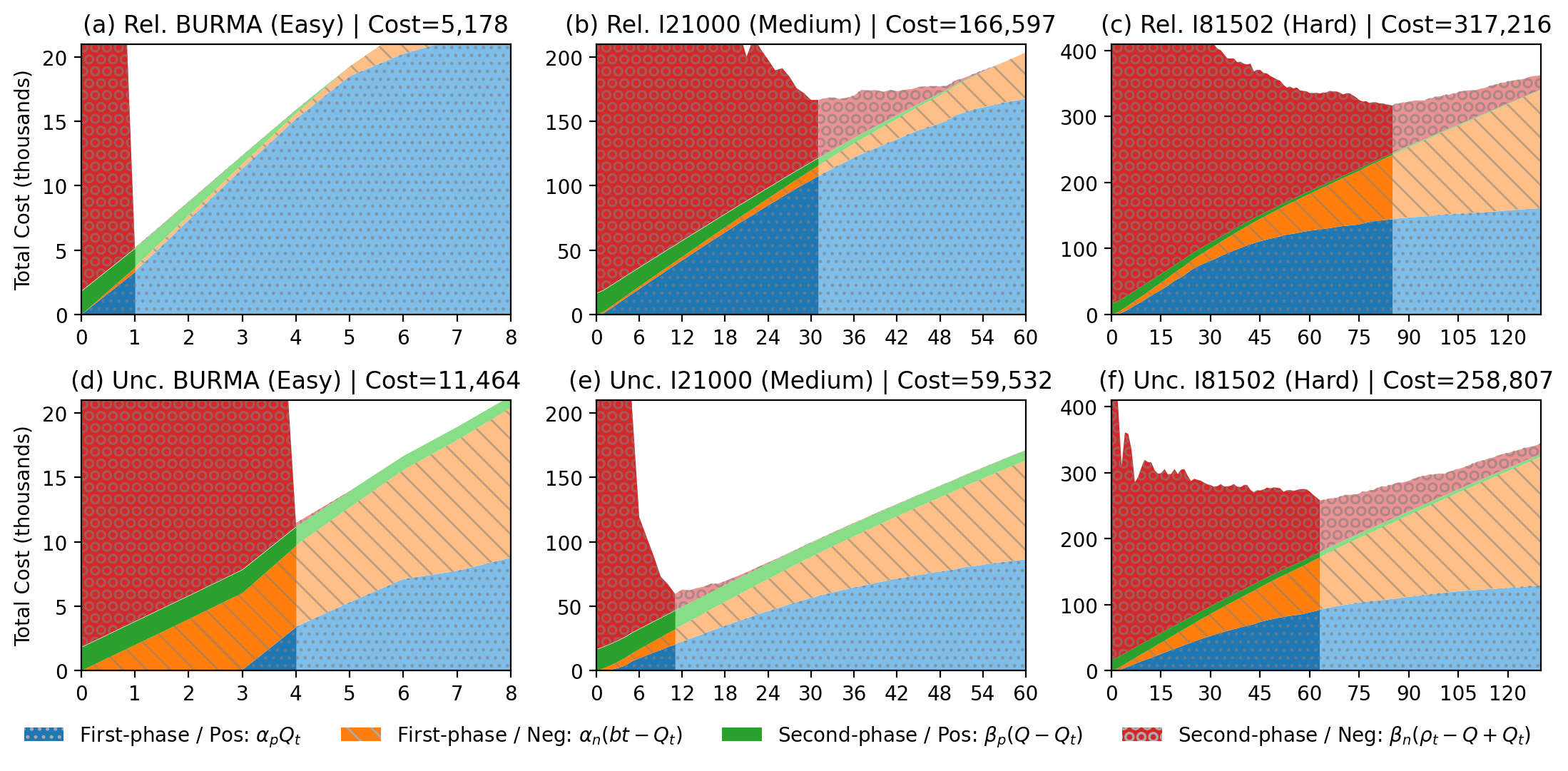}
    \vspace{-1.5em}
    \caption{Cost dynamics graphs exemplifying the interaction of active learning method and category properties for cost structure $(20, 10, 2, 1)$. BURMA is in category bin Easy/Rare, I21000 is in Medium/Common, and I81502 is in Hard/Common. Each graph corresponds to one run (a particular seed), not an average over runs.}
    \label{fig:cost-dynamic-diff-compare}
    \vspace{-1em}
\end{figure*}

When there are few positive examples, relevance feedback and uncertainty sampling act similarly~\cite{lewis1995sequential}. Conversely, when faced with many easily detected positive examples, relevance feedback can drown in non-informative positive examples. 

Table~\ref{tab:breakdown-cost-of-2p} examines this effect, comparing uncertainty sampling and relevance feedback for two-phase workflows on the full and 20\% RCV1-v2 collections, while also partitioning results by category prevalence and difficulty.

\subsubsection{Number of Target Documents}

Both higher category prevalence and larger collection size increase the number of positive examples a workflow must find to hit a recall target. Both therefore provide more opportunity to amortize the cost of any negative examples seen during training.  

Table~\ref{tab:breakdown-cost-of-2p} shows this effect is powerful: in all 36 scenarios examined (3 difficulty levels, 6 cost structures, 2 collection sizes), uncertainty sampling has a bigger advantage (or smaller disadvantage) vs. relevance feedback for Common categories than for Rare ones. Further, in 50 out of 54 comparisons (3 difficulties, 3 prevalences, 6 cost structures), uncertainty sampling improves versus relevance feedback when moving from 20\% RCV1-v2 to the full RCV1-v2.

The effect is particularly strong in expensive training scenarios. For example, on Medium/Common categories with cost structure $(1, 1, 1, 1)$ uncertainty sampling shows a small improvement (from -0.0546 to 0.0200) when going from 20\% RCV1-v2 to full RCV1-v2. With cost structure $(10,10,1,1)$ the improvement is much larger: from 0.0836 to 0.4626.

\subsubsection{Task Difficulty}

The story is less straightforward for task difficulty. If the boundary between positive and negative documents is complex but learnable, techniques like uncertainty sampling can choose informative negative examples. On the other hand, good classifier effectiveness is simply impossible for some tasks, due to noisy labeling or model limitations.  In those cases, relevance feedback might dominate by prioritizing the least bad predictions.  

Table~\ref{tab:breakdown-cost-of-2p} shows that in 33 of 36 scenarios (3 prevalences, 6 cost structures, 2 collection sizes) uncertainty sampling does better relative to relevance feedback on Hard categories than Easy ones. This supports the suggestion that focusing training on classifier effectiveness is desirable on difficult tasks.

The worst tasks for uncertainty sampling are Easy/Rare categories. One can quickly get a good classifier, so reviewing negative examples is almost a pure loss. One should heavily weight exploitation over exploration \cite{osugi2005balancing}. 

We display this for one run (using a seed chosen for typical behavior) of Easy/Rare category \texttt{BURMA} with cost structure (20,10,2,1) in Figures~\ref{fig:cost-dynamic-diff-compare}(a) and (d).  Optimal effectiveness with relevance feedback comes from stopping after a single training batch of size 200, having reviewed almost no negative examples, and immediately exploiting the classifier in phase two. Uncertainty sampling must absorb many negative phase one examples (orange w/ strokes) before finding enough positive examples for switching to phase two to be optimal.

Conversely, Table~\ref{tab:breakdown-cost-of-2p} shows that uncertainty sampling is strongly dominant on Medium/Common tasks, particularly when training is expensive.  
Figures~\ref{fig:cost-dynamic-diff-compare}(b) and (e) show Medium/Common task \texttt{I21000} (Metal Ore Extraction). By selecting a balanced set of positives (blue/dots) and negatives (orange/strokes) examples during training, uncertainty sampling optimally stops at iteration 11, deploying an effective but imperfect classifier in the low cost second phase. Relevance feedback gorges on high cost, low value positive examples during training, with its best case corresponding to deploying a bad classifier at iteration 31. 

For difficult tasks, all approaches are expensive, but the averaged results in Table~\ref{tab:breakdown-cost-of-2p} show uncertainty sampling with a modest advantage.  Figures~\ref{fig:cost-dynamic-diff-compare}(c) and (f) show Hard/Common task \texttt{I81502} (Banking and Financial Services).  Obtaining even a mediocre classifier requires a lot of expensive training, and uncertainty sampling does a better job of grinding this out for this high prevalence category. Conversely, 
Table~\ref{tab:breakdown-cost-of-2p} shows relevance feedback is favored for low prevalence difficult categories, where there are fewer positives over which to amortize training effort.

\subsection{Optimal Stopping Iteration}

\begin{figure*}
    \centering
    \includegraphics[width=0.95\linewidth]{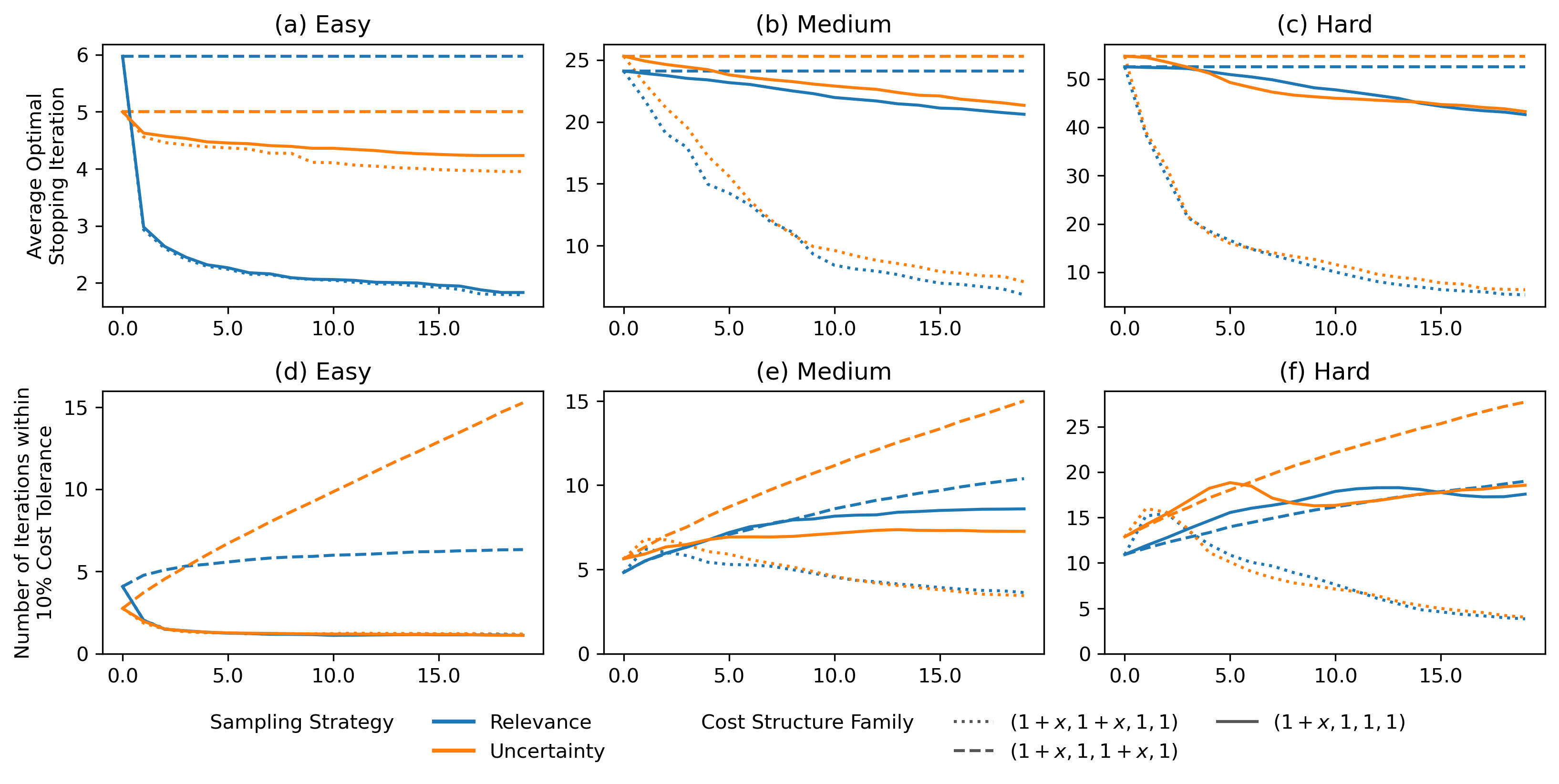}
    \vspace{-1.8em}
    \caption{Mean optimal stopping iteration (top row) and acceptable stopping range with 10\% cost tolerance (bottom row) as asymmetry (value of x on x-axis) is varied within three cost structure families on 20\% RCV1-v2. Means are over 150 runs (15 categories, 10 seeds each).}
    \label{fig:optimal-stopping-comparison}
    \vspace{-1em}
\end{figure*}

Under our idealized cost model, the optimal stopping point is the one that, for a one-phase review, minimizes the sum of review cost and failure cost. Equivalently, this is the optimal point to transition to the second phase of a two-phase review. By providing a precise definition of optimal stopping, our framework allows studying how task properties impact the difficulty faced by stopping rules.  

We examine two characteristics of the stopping problem: the optimal stopping iteration (Figure~\ref{fig:optimal-stopping-comparison}a, b, and c) and the number of iterations during which cost is near-optimal (within 10\% of optimal)  (Figure~\ref{fig:optimal-stopping-comparison}d, e, and f). The smaller the second value, the more difficult the stopping rule's task is.

We average these quantities separately across tasks within the Easy, Medium, and Hard bins. The 20\% RCV1-v2 subset is used to reduce computation time, since many runs must be extended well past the 0.80 recall target to capture the 10\% cost range. We examine how stopping characteristics change as we vary the degree of cost asymmetry within three cost structure families: $(1+x, 1+x, 1, 1)$, $(1+x, 1, 1+x, 1)$, and $(1+x, 1, 1, 1)$ with adjustable $x$ from 0.0 to 20.0.

For the Expensive Training cost family $(1+x, 1+x, 1, 1)$ we see, unsurprisingly, that the optimal stopping iteration decreases quickly as $x$ increases.  Less obviously, the number of iterations during which near-optimal cost can be achieved also narrows as $x$ becomes large: the stakes become higher for stopping outside a narrow range of iterations. Unfortunately, the position of that range varies substantially with task difficulty, posing a challenge for stopping rules.

For the Additive Positive cost family $(1+x,1,1+x,1)$, recall that our analysis showed that the optimal stopping point is independent of $x$ (Section~\ref{subsubsec:additivepositives}). Figures~\ref{fig:optimal-stopping-comparison}a, b, and c confirm this empirically. On the other hand, larger $x$ increases the minimum cost and thus increases the range of iterations where total cost is within 10\% of that minimum. This eases the stopping rule's task, particularly for uncertainty sampling (which can amortize its negative examples over a larger total cost).

Finally, to exhibit the complexities possible, we consider the unusual cost family $(1+x, 1,1,1)$: positives incur extra cost only during training. Since it does not matter when negatives are reviewed, the optimal stopping iteration decreases more slowly with $x$ than for $(1+x, 1+x, 1, 1)$. The range of acceptable stopping iterations is relatively stable, but oscillates with $x$ for Hard categories. The sensitivity of $(1+x,1,1,1)$ to the number of positive training documents is a likely contributor to oscillation, but we are not sure if this is a systematic phenomenon or a peculiarity of the small set of Hard categories used.

%% file: 7-summary.tex
\section{Summary and Future Work}

Our proposed TAR cost model that accounts for cost asymmetries observed in real-world applications, both across document types and across phases of review. We show analytically and empirically that these asymmetries impact which choice of workflow and active learning method minimizes total review cost.  One-phase workflows dominate when costs are uniform, while two-phase workflows are favored (providing up to 60\% cost reductions) when costs are asymmetric. We also show that task characteristics interact with these choices in predictable ways, with the ability to amortize training costs across a larger number of sought documents a major factor. 

We also show how the cost structure impacts the optimization problem faced by stopping rules, which may give insight into their design. We hope that our results will also provide practical guidance to legal technology practitioners, and discourage claims of uniform superiority for one workflow or another.